\documentclass[prl,aps,twocolumn,floatfix,showpacs]{revtex4}
\usepackage{graphics}
\begin{document}
\title{Superfluid and insulating phases of fermion mixtures in optical lattices}
\author{M. Iskin and C. A. R. S{\'a} de Melo}
\affiliation{School of Physics, Georgia Institute of Technology, Atlanta, Georgia 30332, USA}
\date{\today}

\begin{abstract}
The ground state phase diagram of fermion mixtures in optical lattices is analyzed 
as a function of interaction strength, fermion filling factor and tunneling parameters.
In addition to standard superfluid, phase-separated or coexisting superfluid/excess-fermion phases 
found in homogeneous or harmonically trapped systems, fermions in optical lattices have
several insulating phases, including a molecular Bose-Mott insulator (BMI), 
a Fermi-Pauli (band) insulator (FPI), a phase-separated BMI/FPI mixture 
or a Bose-Fermi checkerboard (BFC). 
The molecular BMI phase is the fermion mixture counterpart of the
atomic BMI found in atomic Bose systems, the BFC or BMI/FPI phases exist
in Bose-Fermi mixtures, and lastly the FPI phase is particular to the Fermi
nature of the constituent atoms of the mixture.
 
\end{abstract}
\maketitle

Ultra-cold atoms in optical lattices are ideal systems to study novel condensed matter phases.
The study of Bose atoms in optical lattices has revealed (in addition to superfluid phases)
an atomic Bose-Mott insulator (BMI) phase~\cite{greiner-2002}.
Eventhough great success was achieved in cooling and studying Bose atoms in optical traps,
it was not until very recently that Fermi atoms (mixtures of two-hyperfine states)~\cite{mit-lattice}
or mixtures of Bose and Fermi atoms were sucessfully loaded into optical lattices~\cite{bongs-2006}.
In addition, several groups around the world are attempting to create mixtures of two types 
of fermions, as it was achieved with two types of bosons~\cite{inguscio-2002}.

In a very recent paper the MIT group produced preliminary experimental evidence 
for superfluid and insulating phases of ultracold $^6$Li atoms in optical lattices~\cite{mit-lattice}. 
This last experiment overcame some earlier difficulties of producing Fermi 
superfluids from an atomic Fermi gas or from molecules of Fermi 
atoms in optical lattices~\cite{modugno, kohl, stoferle}.
Unlike in homogeneous or harmonically trapped systems, optical lattices offer an enormous degree
of control since phase diagrams may be studied as a function of 
tunneling matrix element $t_\sigma$ between adjacent lattice sites, 
on-site atom-atom interactions $g$, filling fraction $n_{\sigma}$, 
lattice dimensionality $\cal{D}$ and tunneling anisotropy $\eta = t_\downarrow/t_\uparrow$, 
where $\sigma$ labels the type of fermion state. 

The research explosion that followed successful loading of bosons in optical lattices 
and the observation of Bose-Mott phases~\cite{bloch-review} almost 
guarantees {\it a priori} another research explosion following successful loading of 
fermions in optical lattices and the observation of superfluid and
insulating phases~\cite{mit-lattice}, in particular because fermions are more "fundamental" 
particles of atomic and condensed matter systems in the sense
that they can lead to Bose-like behavior (Bose molecules made of two-fermions) and to combined
Bose-Fermi behavior when there are Bose molecules and unbound excess fermions. 
Thus, the resulting quantum phases of Fermi mixtures is much richer than those 
present in systems consisting of atomic bosons or Bose-Fermi mixtures in optical lattices.
Arguably, mixtures of two-hyperfine-states of the same type of fermion or mixtures 
of two different types of fermions loaded into optical lattices are
one of the next frontiers in ultracold atom research because of their greater tunability and 
the richness of their phase diagrams.

Our main results are as follows.
Using an attractive Fermi-Hubbard Hamiltonian to describe fermion mixtures in 
optical lattices, we obtain the ground state phase 
diagram containing normal, phase-separated and coexisting superfluid/excess-fermions, and 
insulating regions as a function of interaction strength and density of fermions.
We show that when fermion-fermion (Bose) molecules are formed they interact with each other
strongly and repulsively. Furthermore, when there are excess fermions, 
the resulting system corresponds to a strongly interacting (repulsive) mixture of bosons and fermions
in the molecular limit, in sharp contrast with homogenous systems where 
the resulting Bose-Fermi mixtures are weakly interacting~\cite{pieri}.
This result is a direct manifestation of the Pauli exclusion principle in the lattice case, 
since each Bose molecule consists of two fermions, and more than one identical 
fermion on the same lattice site is not allowed. 
Lastly, several insulating phases appear in the strong attraction limit depending 
on the fermion filling fractions.
For instance, we find a molecular Bose-Mott insulator (superfluid) when the molecular filling 
fraction is equal to (less than) one when the fermion filling fractions
are identical in qualitative agreement with the MIT experiment~\cite{mit-lattice}.
Furthermore, when the filling fraction of one type of fermion is one and 
the filling fraction of the other is one-half (corresponding to molecular boson 
and excess fermion filling fractions of one-half),
we also find either a phase-separated state consisting of a Fermi-Pauli insulator (FPI) 
of the excess fermions and a molecular Bose-Mott insulator (BMI) or a Bose-Fermi checkerboard 
(BFC) phase depending on the tunneling anisotropy $\eta$. Finally, we propose
that all these superfluid and insulating phases can be observed in mixtures of fermions
loaded into optical lattices.

{\it Lattice Hamiltonian:}
To describe mixtures of fermions loaded into optical lattices, we start with a
single-band Fermi-Hubbard Hamiltonian in momentum space
\begin{eqnarray}
\label{eqn:hamiltonian}
H = \sum_{\mathbf{k},\sigma}\xi_{\mathbf{k},\sigma} a_{\mathbf{k}, \sigma}^\dagger a_{\mathbf{k}, \sigma} 
- \frac{g}{2} \sum_{\mathbf{k},\mathbf{k',\mathbf{q},\sigma}}
b_{\mathbf{k},\mathbf{q},\sigma}^\dagger b_{\mathbf{k'},\mathbf{q},\sigma}, 
\end{eqnarray}
with an on-site attractive interaction $g > 0$.  
Here, the pseudo-spin $\sigma$ labels the trapped hyperfine states of a given species of fermions,
or labels different types of fermions in a two-species mixture, where
$a_{\mathbf{k}, \sigma}^\dagger$ is the creation operator and
$b_{\mathbf{k},\mathbf{q},\sigma}^\dagger = a_{\mathbf{k}+\mathbf{q}/2,\sigma}^\dagger 
a_{-\mathbf{k}+\mathbf{q}/2,-\sigma}^\dagger$.
In addition,  
$
\xi_{\mathbf{k},\sigma}= \epsilon_{\mathbf{k},\sigma} - \widetilde{\mu}_\sigma
$ 
describes the nearest neighbor tight-binding dispersion
$
\epsilon_{\mathbf{k},\sigma} = 2t_\sigma \theta_{\mathbf{k}}
$ 
with $\widetilde{\mu}_\sigma = \mu_\sigma - V_{H,\sigma}$
and $\theta_{\mathbf{k}} = \sum_{i} [1 - \cos (k_i a_c)]$,
where $t_\sigma$ is the tunelling matrix element, $\mu_\sigma$ is the chemical potential,
$V_{H,\sigma}$ is a possible Hartree energy shift and $a_c$ is the lattice spacing.
Notice that, we allow fermions to be of different species through $t_\sigma$, and
to have different populations controlled by independent $\widetilde{\mu}_\sigma$.
Furthermore, unlike recent work of BCS pairing of fermions in optical lattices~\cite{liu-lattice,torma-lattice},
we discuss the evolution from BCS to BEC pairing and the emergence of insulating phases.
We ignore multi-band effects since a single-band Hamiltonian 
may be sufficient to describe the evolution from BCS to BEC physics
in optical lattices~\cite{stoof-2006,footnote1}. However, these effects can be easily incorporated
into our theory.

For the Hamiltonian given in Eq.~(\ref{eqn:hamiltonian}), 
the saddle point order parameter equation is given by
\begin{equation}
\frac{1}{g} = \frac{1}{M}\sum_{\mathbf{k}} \frac{1 - f(E_{\mathbf{k},\uparrow}) - f(E_{\mathbf{k},\downarrow})} 
{2E_{\mathbf{k},+}},
\label{eqn:op}
\end{equation}
where $M$ is the number of lattice sites,
$
f(x) = 1/[\exp(x/T) + 1]
$
is the Fermi function,
$
E_{\mathbf{k},\sigma} = (\xi_{\mathbf{k},+}^2 + |\Delta_0|^2)^{1/2} + s_\sigma \xi_{\mathbf{k},-}
$
is the quasiparticle energy when $s_\uparrow = 1$ or
the negative of the quasihole energy when $s_\downarrow = -1$, and
$
E_{\mathbf{k},\pm} = (E_{\mathbf{k},\uparrow} \pm E_{\mathbf{k},\downarrow})/2.
$
Here, $\Delta_0$ is the order parameter and
$
\xi_{\mathbf{k},\pm} = \epsilon_{\mathbf{k},\pm} - \widetilde{\mu}_\pm,
$
where 
$
\epsilon_{\mathbf{k},\pm} = 2t_\pm \theta_{\mathbf{k}}
$
with $t_\pm = (t_\uparrow \pm t_\downarrow)/2$ and 
$
\widetilde{\mu}_\pm = (\widetilde{\mu}_\uparrow \pm \widetilde{\mu}_\downarrow)/2.
$
Notice that, the symmetry between quasiparticles and quasiholes 
is broken when $\xi_{\mathbf{k},-} \ne 0$. 
The order parameter equation has to be solved self-consistently with number equations
\begin{eqnarray}
N_{\sigma} = \sum_{\mathbf{k}} 
\left[ |u_{\mathbf{k}}|^2 f(E_{\mathbf{k},\sigma})+ |v_{\mathbf{k}}|^2 f(-E_{\mathbf{k},-\sigma}) \right],
\label{eqn:number}
\end{eqnarray}
where 
$
|u_{\mathbf{k}}|^2 = (1 + \xi_{\mathbf{k}, +}/E_{\mathbf{k}, +})/2
$
and
$
|v_{\mathbf{k}}|^2 = (1 - \xi_{\mathbf{k}, +}/E_{\mathbf{k}, +})/2.
$
The number of $\sigma$-type fermions per lattice site is given by 
$0 \le n_\sigma = N_\sigma/M \le 1$. Thus, when
$n_\uparrow \ne n_\downarrow$, we need to solve all
three self-consistency equations, since population imbalance is achieved
when either $E_{\mathbf{k},\uparrow}$ or $E_{\mathbf{k},\downarrow}$ is negative 
in some regions of momentum space, as discussed next.

\begin{figure} [htb]
\centerline{\scalebox{0.38}{\hskip -11mm \includegraphics{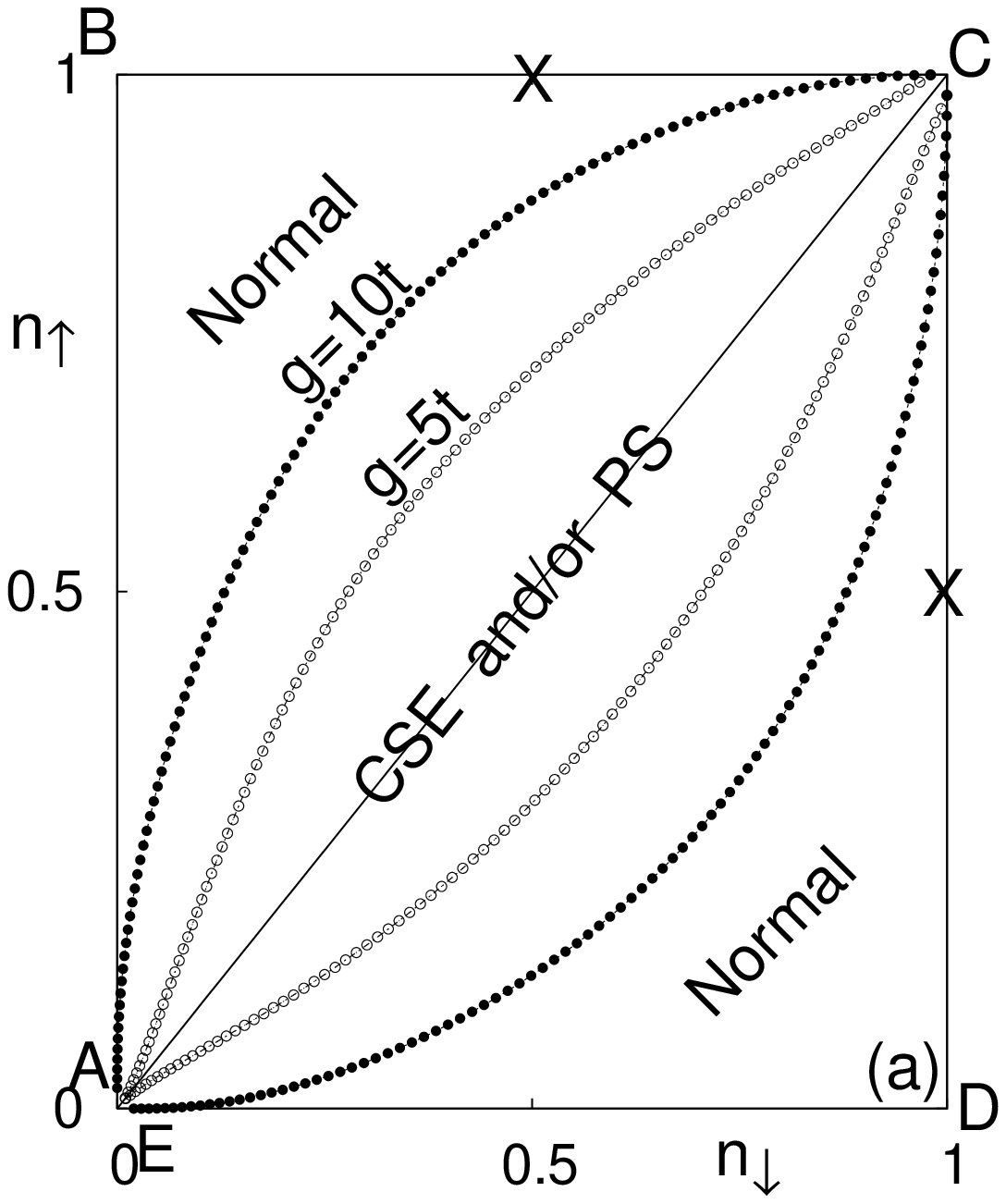} \hskip -15mm \includegraphics{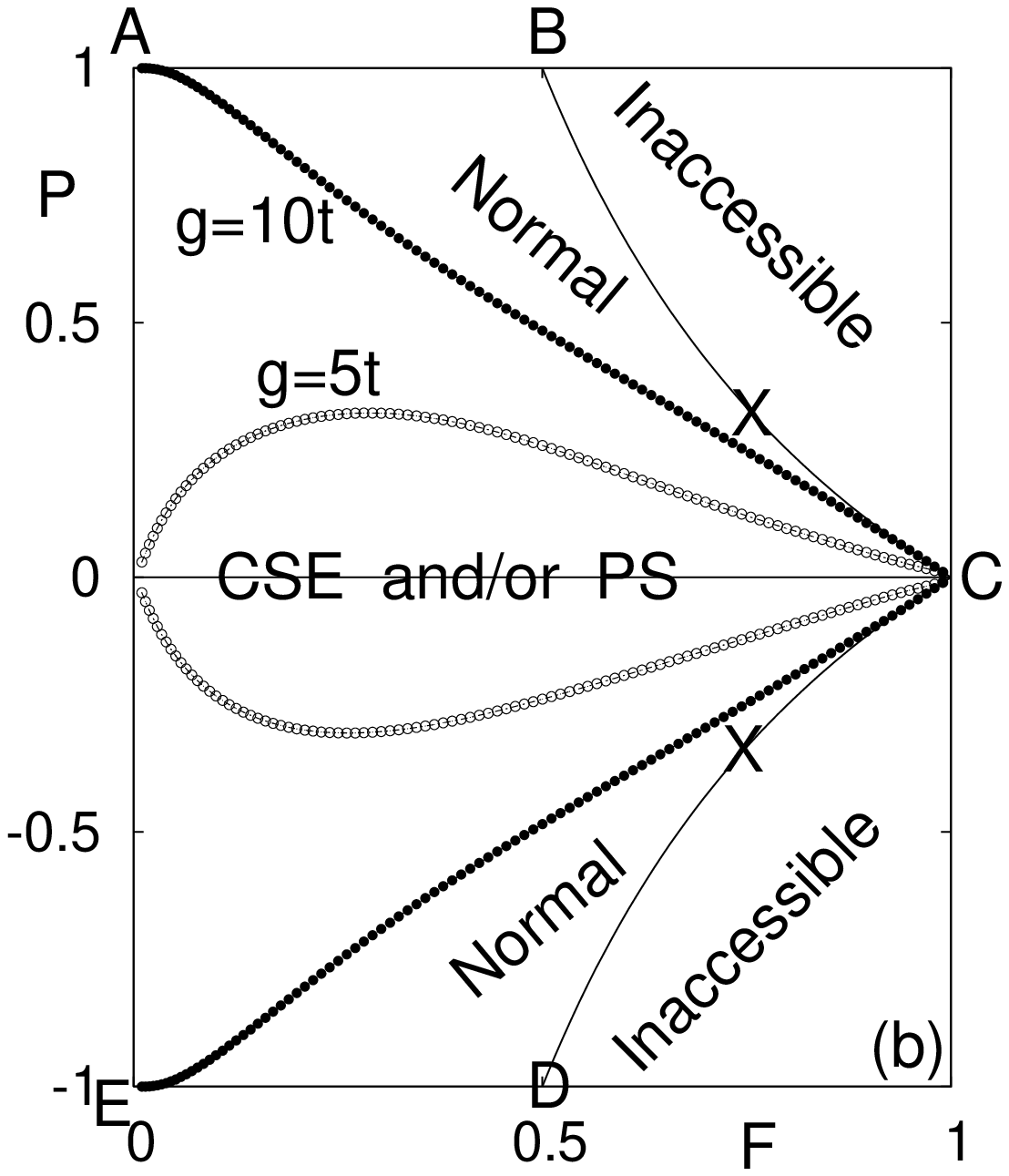}}}
\caption{\label{fig:phase1}
(a) $n_\uparrow$ versus $n_\downarrow$, and
(b) $P$ versus $F$ diagrams
for $g = 5t$ and $g = 10t$. The normal regions (outside the ``football'') 
and coexistence of superfluidity with excess fermions (CSE) and/or 
phase separation (PS) (inside the ``football'') are indicated. 
The CSE/PS (normal) region expands (shrinks) with increasing attraction. 
}
\end{figure}

{\it Ground state saddle point phase diagrams:}
To obtain ground state phase diagrams, 
we solve Eqs.~(\ref{eqn:op}) and~(\ref{eqn:number})
as a function of interaction strength $g$, 
population imbalance $-1 \le P = (n_\uparrow - n_\downarrow)/(n_\uparrow + n_\downarrow) \le 1$ and
total filling fraction $0 \le F = (n_\uparrow + n_\downarrow)/2 \le 1$,
for two sets of tunneling ratios $\eta = t_\downarrow / t_\uparrow$.
The case of $\eta = 1$ ($t_\sigma = t$) is shown in Fig.~\ref{fig:phase1}, and 
the case of $\eta = 0.15$ is not shown.
While $\eta = 1$ corresponds to one-species (two-hyperfine-state) mixture such
as $^6$Li or $^{40}$K, $\eta = 0.15$ corresponds to a two-species mixture (one-hyperfine-state of each
type of atom) such as $^6$Li and $^{40}$K.

In the phase diagrams shown in Fig.~\ref{fig:phase1}, 
we indicate the regions of normal (N) phase where $|\Delta_0| = 0$, and group
together the regions of coexistence of superfluidity and excess fermions (CSE) 
and/or phase separation (PS), where $|\Delta_0| \ne 0$.
When $F \ll 1$, the phase diagrams are similar to the homogenous case~\cite{pao, sheehy}, 
and the $P$ versus $F$ phase diagram is symmetric for 
equal tunnelings as shown in Fig.~\ref{fig:phase1}, and 
is asymmetric for unequal tunnelings having a smaller normal region 
when the lighter band mass fermions are in excess (not shown)~\cite{footnote2}. 
Here, we do not discuss separately the CSE and PS regions since they have already been 
discussed in homogeneous and harmonically trapped systems~\cite{pao, sheehy}, and
experimentally observed~\cite{mit, rice}, but we make two remarks.
First, the phase diagram characterized by normal, non-normal (CSE or PS), and insulating 
regions may be explored experimentally by tuning the ratio $g/t_+$, total filling fraction $F$, 
and population imbalance $P$ as done in harmonic traps~\cite{mit, rice}.
Second, topological phases characterized by the number (I and II) of simply connected 
zero-energy surfaces of $E_{\mathbf{k},\sigma}$ may lie in the stable region of CSE,
unlike in the homogeneous case where the topological phase II 
always lies in the phase separated region for all parameter space~\cite{iskin-mixture}. 

We would like to emphasize that the saddle point approximation only tells us
that the system is either superfluid $(\vert \Delta_0 \vert \ne 0)$ or normal $(\vert \Delta_0 \vert = 0 )$, but 
fails to tells us about insulating phases. Thus, first, we present a physical discussion
of the emergence of insulating phases, and then show that these phases indeed emerge from 
fluctuation effects beyond the saddle point approximation.

{\it Emergence of Insulating Phases.} Generally, lines AB ($0 < n_\uparrow < 1$; $n_{\downarrow} = 0$) and 
ED ($n_\uparrow = 0$; $ 0 < n_{\downarrow} < 1$) in Fig.~\ref{fig:phase1} 
correspond to normal $\sigma$-type Fermi gases for all interactions, 
while points B $(n_\uparrow = 1, n_\downarrow = 0)$ and D $(n_\uparrow = 0, n_\downarrow = 1)$
correspond to a Fermi-Pauli (band) insulator since there is only one type of
fermion in a fully occupied band. Thus, the only option for additional 
fermions $(\uparrow$ in case B and $\downarrow$ in case D) 
is to start filling higher energy bands if the optical potential supports it, 
otherwise the extra fermions are not trapped. For the case where no additional bands are occupied we
label the corresponding phase diagram regions as `Inaccessible' 
in Fig.~\ref{fig:phase1}(b), since either $n_\uparrow > 1$ or $n_\downarrow > 1$
in these regions.

In addition, the population balanced line AC ends at the special point C, 
where $n_\uparrow = n _\downarrow = 1$. This point is a Fermi-Pauli (band) insulator
for weak attraction since both fermion bands are fully occupied, and a Bose-Mott Insulator (BMI) 
in the strong attraction limit, since at each lattice site 
there is exactly one molecular boson (consisting of a pair of $\uparrow$ and $\downarrow$ fermions) 
which has a strong repulsive on-site interaction 
with any additional molecular boson due to the Pauli exclusion principle.

\begin{figure} [htb]
\centerline{\scalebox{0.3}{\includegraphics{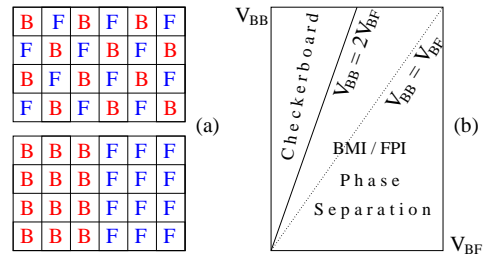}} }
\caption{\label{fig:phase3}
(Color online) (a) schematic diagram for the BFC phase (top) 
and BMI/FPI phase separation (bottom), and 
(b) $V_{BB}$ versus $V_{BF}$ phase diagram. 
}
\end{figure}

Furthermore, for very weak fermion attraction, lines 
BC ($n_\uparrow = 1, 0 < n_\downarrow < 1$) and CD 
($0 < n_\uparrow < 1, n_\downarrow = 1$) correspond essentially 
to a fully polarized ferromagnetic metal (or half-metal), 
where only the fermion with filling fraction less than one can move around.   
However, when the fermion attraction is sufficiently strong the
lines BC and CD must describe insulators, as molecular bosons
and excess fermions are strongly repulsive due to the Pauli exclusion 
principle. The crosses in Fig.~\ref{fig:phase1} at points $n_\uparrow = 1, n_\downarrow = 1/2$
or $n_\uparrow = 1/2, n_\downarrow = 1$ indicate the case where 
the molecular boson filling fraction 
$n_B = 1/2$ and the excess fermion filling fraction is $n_e = 1/2$.
At these high symmetry points, molecular bosons and excess fermions tend
to segregate, either producing a domain wall type of phase separation with a molecular 
Bose-Mott insulator (BMI) and a Fermi-Pauli insulator (FPI) region 
or a checkerboard phase of alternating molecular bosons and excess fermions (BFC).
A schematic diagram of these two phases is shown in Fig.~\ref{fig:phase3}(a).

Thus, the strong attraction limit in optical lattices brings additional physics 
not captured at the saddle point and not present in homogenous or purely harmonically 
trapped systems, as discussed next.

{\it Strong attraction (molecular) limit:} The emergence of insulating phases in the
strong attraction attraction limit requires the simultaneous inclusion of spatial and
temporal fluctuations. Thus, first, we derive a time dependent Ginzburg-Landau theory
involving molecular bosons and excess fermions near the critical temperature $T_c$ 
of the possible superfluid phase leading to
$
[ a + b|\Lambda(x)|^2 - c\nabla^2/2 - id(\partial / \partial t) ] \Lambda(x) = 0
$
in the $x = (\mathbf{x},t)$ representation.
Here, $\Lambda(x)$ is the fluctuation of the order parameter around its saddle point value $|\Delta_0| = 0$.

In the strong attraction (molecular) limit
$|\widetilde{\mu}_+| \approx |\epsilon_b|(1 - p_e)/2 \gg 2{\cal D}t_+$, we obtain
$a = a_1 + a_2 = -[2\widetilde{\mu}_+ - \epsilon_b(1-p_e)]/[g^2(1 - p_e)] + p_e/[g(1-p_e)]$,
$b = b_1 + b_2 = 2/[g^3(1-p_e)^2] - (\partial p_e/\partial \widetilde{\mu}_e)/[g^2(1-p_e)]$,
$c = 4a_c^2 t_\uparrow t_\downarrow/[g^3(1-p_e)^2]$, and
$d = 1/[g^2(1-p_e)]$.
Here, $\epsilon_b = - g$ is the binding energy defined by
$
1/g = \sum_{\mathbf{k}}1/(2\epsilon_{\mathbf{k},+} - \epsilon_b),
$
and $e$ ($-e$) labels the excess (non-excess) type of fermions and 
$p_e = | n_{\uparrow} - n_{\downarrow}| $ is the number of unpaired fermions per lattice site.

Through the rescaling $\Psi(x) = \sqrt{d}\Lambda(x)$,
we obtain the equation of motion for a mixture of bound pairs (molecular bosons)
and unpaired (excess) fermions
\begin{eqnarray}
-\mu_B \Psi(x) &+& \left[U_{BB}|\Psi(x)|^2 + U_{BF} p_e(x) \right] \Psi(x) \nonumber \\ 
&-& \frac{\nabla^2 \Psi(x)}{2m_B}  - i\frac{\partial \Psi(x)}{\partial t} = 0,
\label{eqn:GL-BEC}
\end{eqnarray}
with pair chemical potential 
$
\mu_B = - a_1/d = 2\widetilde{\mu}_+ - \epsilon_b(1-p_e),
$
mass 
$
m_B = d/c = g/(4a_c^2 t_\uparrow t_\downarrow),
$
and repulsive pair-pair 
$
U_{BB} = b_1 a_c^3/d^2 = 2g a_c^3
$
and pair-fermion
$
U_{BF} = a_2 a_c^3 /(d p_e) = g a_c^3
$
interactions.
This procedure also yields
$
p_e(x) = [a_2/d + b_2|\Psi(x)|^2/d^2]/U_{BF} 
= p_e - g a_c^3(\partial p_e/\partial \mu_e) (1-p_e) |\Psi(x)|^2 \ge 0,
$
which is the spatial density of unpaired fermions. In contrast with 
the homogeneous or harmonically trapped systems~\cite{pieri, iskin-mixture}, the boson-boson and
boson-fermion interactions are strongly repulsive due to the important role
played by the Pauli exclusion principle in the lattice, which
is discussed next.

{\it Effective lattice Bose-Fermi action:} In the limit of strong attractions between
fermions $g/t_{+} \gg 1$, we obtain an effective Bose-Fermi lattice action 
\begin{equation}
S_{BF} =  \int_0^{\beta} d\tau  \left[ \sum_{i} ( f_i^\dagger \partial_{\tau} f_i + 
b_i^\dagger \partial_{\tau} b_i)  +  {H}_{BF}^{eff} \right], 
\end{equation}
where
$
{H}_{BF}^{eff} = K_{F} + K_{B} + H_{BF} + H_{BB}.
$
Here, $K_{F} = -\mu_F \sum_i  f_i^\dagger f_i 
- t_F \sum_{\langle i,j \rangle} f_i^\dagger f_j $ 
is the kinetic part of the excess fermions;
$K_B = - \mu_{B} \sum_i  b_i^\dagger b_i 
- t_B \sum_{\langle i,j \rangle} b_i^\dagger b_j$
is the kinetic part of the molecular bosons;
$H_{BF} =  U_{BF} \sum_{i} f_i^\dagger f_i b_i^\dagger b_i$
is the interaction between molecular bosons and excess
fermions; and $H_{BB} = U_{BB} \sum_{i} b_i^\dagger b_i b_i^\dagger b_i $
is the interaction between two molecular bosons.

The total number of fermions is fixed by the constraint
$n = 2n_B + p_e$, where $n_B = N_B/M$ is the number of bosons per lattice site.
The important parameters of this effective Hamiltonian are the excess fermion transfer energy $t_F = t_e$, 
the molecular boson transfer energy $t_B = 2 t_{\uparrow} t_{\downarrow}/g$,
the boson-fermion effective repulsion $U_{BF} = g$ and the boson-boson
effective repulsion $U_{BB} = 2g$.
Notice that, on-site interactions $U_{BB}$ and $U_{BF}$ become infinite (hard-core)
when $g \to \infty$ as a manifestation of the Pauli exclusion principle.
In addition, there are weak and repulsive nearest neighbor boson-boson 
$V_{BB} \propto (t_{\uparrow}^2 + t_{\downarrow}^2)/g$ and
boson-fermion $V_{BF} \propto t_e^2/g$ interactions. 
These repulsive interactions in optical lattices lead to several insulating phases,
depending on fermion filling fractions.
In the following analysis, we discuss only two high symmetry cases:
(a) $n_\uparrow = n_\downarrow$; and
(b) $n_\uparrow = 1$ and $n_\downarrow = 1/2$ or  $n_\uparrow  = 1/2$ and $n_\downarrow = 1$.

\begin{figure} [htb]
\centerline{\scalebox{0.3}{\includegraphics{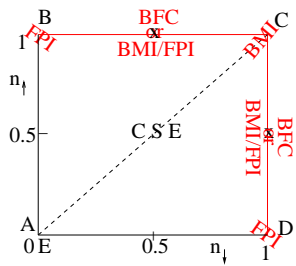}} }
\caption{\label{fig:phasei}
(Color online) $n_\uparrow$ versus $n_\downarrow$ phase diagram 
in the strong attraction limit, indicating the CSE
(superfluid), the metallic (lines AB and DE), and 
insulating (lines BC and CD) phases with special points FPI, BFC, BMI/FPI, and BMI.
}
\end{figure}

In case (a) indicated as point C in Figs.~\ref{fig:phase1} and~\ref{fig:phasei}
where $p_e = 0$, $H_{BF}$ reduces to a molecular Bose-Hubbard Hamiltonian with the 
molecular Bose filling fraction $n_B = n/2 = F$, thus leading to 
a molecular BMI when $n_B = 1$ beyond a critical value of $U_{BB}$.
The critical value $U_{BB}^c$ needed 
to attain the BMI phase can be estimated using the approach of Ref.~\cite{stoof-2001}
leading to $U_{BB}^c = 3 (3 + \sqrt{8}) t_B$, which in terms of the underlying
fermion parameters leads to $g_c = 4.18 \sqrt{t_{\uparrow} t_{\downarrow}}$
for the critical fermion interaction. This value of $g_c$ is just a lower bound
of the superfluid-to-insulator (SI) transition, since $H_{BF}^{eff}$ is only valid in 
the $g \gg t_+$ limit. This SI transition at $g_c$ has been observed 
in recent experiments~\cite{mit-lattice}. 

In case (b) indicated as crosses in Figs.~\ref{fig:phase1} and~\ref{fig:phasei}, 
the ground state of the effective molecular-boson/excess-fermion system corresponds to either 
a checkerboard phase of alternating bosons and fermions
or to a phase separated BMI/FPI system depending on the ratio $V_{BB}/V_{BF}$. 
The checkerboard phase shown in Fig~\ref{fig:phase3}(a) is favored when $V_{BB} > 2V_{BF}$, leading to 
the phase diagram of Fig.~\ref{fig:phase3}(b). At the current level of approximation,
we find that when $t_\uparrow = t_\downarrow$ phase separation is 
always favored, however when $\uparrow$ ($\downarrow$) fermions are in excess the checkerboard phase 
is favored when $t_\downarrow > \sqrt{3} t_\uparrow$ 
$(t_\downarrow < t_\uparrow/\sqrt{3})$.
Therefore, phase separation and checkerboard phases are achievable 
if the tunneling ratio $\eta$ can be controlled experimentally in optical
lattices. Notice that this checkerboard phase present in the lattice case 
is completely absent in homogeneous or harmonically trapped systems~\cite{pao,sheehy,iskin-mixture}.
Furthermore, the entire lines BC and CD in Fig.~3 represent insulating phases.

{\it Conclusions:}
We have analyzed the ground state phase diagram of fermion mixtures in optical lattices as a
function of interaction strength, fermion filling factor, and tunneling parameters.
In addition to standard superfluid, 
phase separated or coexisting superfluid/excess fermion phases, 
we have found several insulating phases including a
molecular Bose-Mott insulator (BMI), a Fermi-Pauli (band) insulator (FPI), a phase separated
BMI/FPI mixture, and a Bose-Fermi checkerboard phase depending 
on fermion filling fractions. All these additional phases make the physics
of Fermi mixtures much richer than those of atomic bosons
or Bose-Fermi mixtures in optical lattices, and of harmonically trapped fermions.
Lastly, the molecular BMI phase discussed here has been 
preliminarily observed in a very recent MIT experiment~\cite{mit-lattice}, 
opening up the experimental exploration of the rich phase diagram of  
fermion mixtures in optical lattices in the near future.
We thank NSF (DMR-0304380) for support.


\begin{thebibliography}{99}
\bibitem{footnote1} Many bands are important when the fermion filling fraction
is higher than one and/or the optical lattice is not in the tight-binding regime.
\end{thebibliography}
\end{document}